# THE MORPHOLOGICAL EVOLUTION OF GALAXIES

RICHARD S ELLIS

Institute of Astronomy

Madingley Road, Cambridge CB3 0HA, UK

August 3, 1995

I review the general progress made in the study of galaxy evolution concentrating on the impact of systematic ground-based spectroscopic surveys of faint galaxies and high resolution imaging with Hubble Space Telescope. The picture emerging is one where massive regular galaxies have changed little since redshift $z \simeq 1$, whereas there has been a marked decline in the abundance of less massive star forming dwarf galaxies. Reconciling these trends with the conventional hierarchical growth of structure on galactic scales and determining the fate of the star forming dwarfs remains a major unsolved problem in contemporary cosmology.

## 1 Introduction

Considerable progress has been made in recent years in constraining the recent evolution of normal galaxies through detailed studies of stellar populations of nearby systems and comprehensive photometric and spectroscopic surveys of field and cluster galaxies viewed at large look-back times. An outstanding restriction in the analysis of high redshift data has been the 'comparing apples and oranges' problem. Given the diverse nature of the galaxy population at any redshift, how can similar types be determined unambiguously over a range in look-back time and the evolution of a subset of the population be reliably determined? Let me start by discussing two recent examples of observational progress in the high redshift area which, as a 'Devil's advocate', I will expose to illustrate some weaknesses we seek to eliminate.

In the first example, let us consider the work of Dressler & Gunn (1990) and Aragón-Salamanca et al (1993) on the respective $D_{4000}$ spectral discontinuities and $V - K$ colours for the reddest galaxies in a number of $z > 0.5$ clusters. By comparing the mean properties of such galaxies with the spectral energy distributions of early-type galaxies observed locally, both teams claim there has been only modest evolution in luminous early-type cluster members since $z \simeq 1$.

Whilst it might seem reasonable to suppose that the *reddest* galaxies in each high $z$ cluster are early-type systems, without morphological imaging one cannot exclude the possibility that some of the *bluer* galaxies are also early-type. The simple conclusion derived by Dressler, Aragón-Salamanca and co-workers only truly applies if the early-type population behaves uniformly so that observations of a few examples, selected by colour or spectral class, are representative of the entire population. As there is no *a priori* reason to suppose the local uniformity in the colour-spectral class-morphology plane extends to high redshift, the absence of any one of these observables could seriously undermine any evolutionary tests.

In the second example, we consider spectroscopic surveys of field galaxies (Broadhurst et al 1988, Colless et al 1990, Cowie et al 1992, Glazebrook et al 1995a) which have collectively reinforced the view that the excess galaxies first noted in the deep photographic and CCD galaxy counts are caused by an increase in the absolute normalisation of the field galaxy luminosity function (LF) with redshift. Some subset of the galaxy population is either disappearing or fading with time. How can examples of this population be located so that their evolution can be tracked independently and studied? Notwithstanding the large investment in redshift surveys, as LFs are statistical by nature, some 'marker' is required to identify the evolving subset.

In this lecture I review the progress in these and related areas that is now possible via the recovery of the originally-anticipated imaging capabilities of Hubble Space Telescope. Morphological classifications are now feasible for remote galaxies at cosmologically interesting redshifts. The morphology of a galaxy is, of course, only some form of visual label with as yet no agreed physical basis. However, I suggest it does provide a good marker for resolving evolutionary impasses such as those discussed above. Conventional photometric and spectroscopic indicators of galaxy class are unstable to sudden changes in the star formation rate and thus it is difficult to use these criteria to reliably connect similar classes across ranges in look-back time. Is a faint blue emission line source the predecessor of a local red quiescent galaxy or new class of object? Although morphologies have disadvantages too (they may also evolve and are difficult to measure at faint limits), the additional dimension is a tremendous bonus in understanding the origin of the Hubble sequence.

## 2 Early Formation of Massive Ellipticals

The repair of HST permits the direct identification of elliptical and S0 galaxies at significant cosmic depths. Figure 1 shows a comparison of two images of the distant cluster 0016+16 ($z$=0.54) taken with similar exposure times and filters using the 4.2m William Herschel telescope in 0.9 arcsec seeing (Smail et al 1994) and WFPC-2 (as part of the 'Morphs' collaboration of Butcher, Couch, Dressler, Ellis, Oemler, Sharples & Smail). The improvement in the ability to classify faint galaxies is evident.

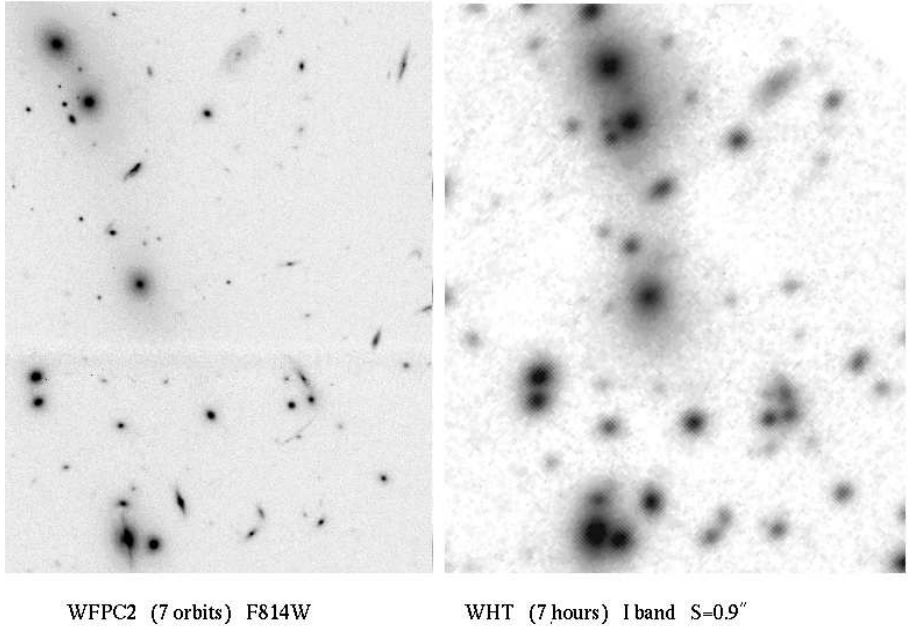

WFPC2 (7 orbits) F814W    WHT (7 hours) I band S=0.9″

Figure 1: *0016+16 z=0.54 as viewed by WFPC-2 (left panel, courtesy of the 'Morphs') and the 4.2m WHT in 0.9 arcsec seeing (right panel, from Smail et al 1994).*

To illustrate the possibilities further, Figure 2 shows a selection of early-type members in 0016+16 drawn from the comprehensive spectroscopic work of Dressler & Gunn (1992); the separation between S0s and ellipticals is apparent. Although inevitably there are some cases where precise classifications remain uncertain, the data offers a number of important possibilities which the Morphs team is exploring.

A particularly interesting, although preliminary, result concerns the photometric homogeneity of the giant elliptical population at these large look-back times. Bower et al (1992) demonstrated, using precision $UBV$ photometry of ellipticals in the Coma and Virgo clusters, that the intrinsic scatter around the $U-V$ colour-luminosity relation is barely detectable ($\delta(U-V)_o$ <0.033). As the $U$-band light is a strong constraint on past star-formation rates (SFR), either ellipticals across both clusters shared the same synchronous decline in SFR (a 'divine intervention' hypothesis), or the homogeneity indicates the systems are sufficiently old for individual variations in the time of formation to be negligible. Depending on the degree of synchronicity, Bower et al claimed ages of between 8 and 12 Gyr. This simple test is remarkably powerful because the constraint depends only on the rate at which the rest-frame $U$-band luminosity declines with time which, in turn, is governed by main sequence lifetimes which are well-understood, at least on a relative scale.

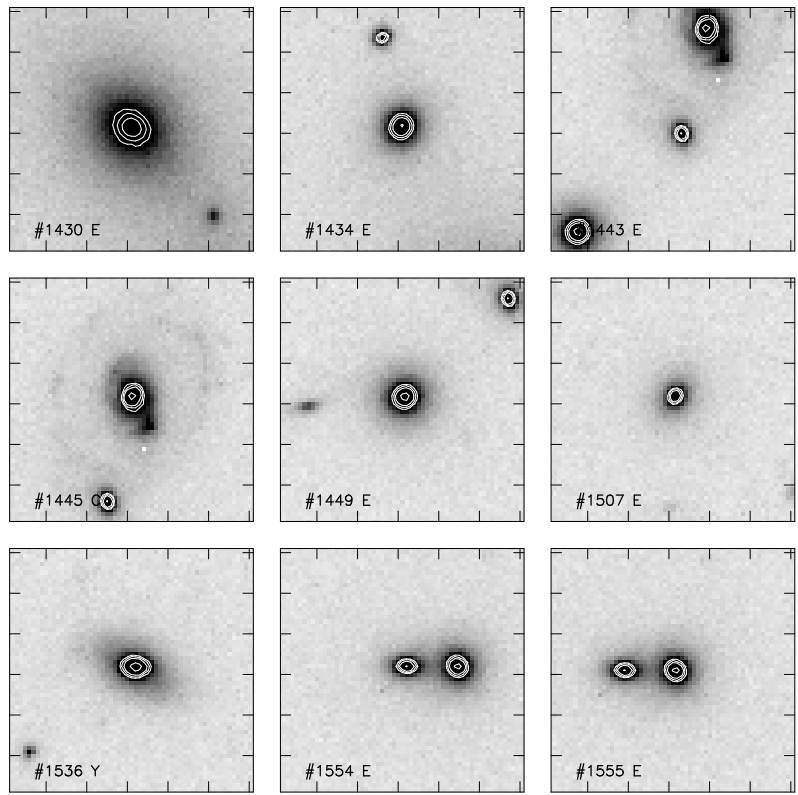

Figure 2: A montage of spectroscopically-confirmed early-type members of 0016+16 (z=0.54) from the 'Morphs' WFPC-2 programme, including one spiral (center left); each box is 5 arcsec on a side. The classification of early-types, and even the distinction between Es and S0s, becomes a practical proposition.

At $z \simeq 0.55$, by good fortune, the $V - I$ colours from HST map neatly onto rest-frame $U - V$ enabling the same test to be carried out for morphologically-confirmed elliptical members at a significant look-back time. The test has been carried out for several of the clusters in the appropriate redshift range in the Morphs' sample and the scatter is somewhat larger, $\delta(U-v)_o <0.07$, the precise upper limit being determined by the increased photometric uncertainties for such faint sources. Even so, this places a tight constraint on the ages of stars in giant galaxies which is then added to the considerable look-back time.

[A preliminary version of this test was described by Ellis (1992) on the basis of ground-based data alone. Quite apart from the morphological confirmation provided by the improved angular resolution of HST, the WFPC-2 photometry is actually *better* than that of the 4.2m Herschel primarily because the sky background is significantly fainter. It is a misconception that HST is a 'small aperture telescope'. For background-limited imaging the reduced background in the near-infrared renders HST equivalent to a 6.5m telescope independent of its superlative angular resolution.]

Depending on the cosmological model, $\delta(U-V)_o$ <0.07 at $z$=0.55, when added to the look-back time, implies a *present-day* age of >12 Gyr. Cluster ellipticals thus join globular clusters in presenting a problem for certain cosmological models. Indeed, there is evidence that regular red ellipticals are seen in abundance in HST images of clusters beyond $z \simeq 1$ (Dickinson 1995), although spectroscopic work is needed in these fields to be sure of the implications.

These tests make strong statements about the ages of galaxies without directly observing their epoch of formation. A good indication of the likely impact of 8-10 m class telescopes in this area follows from the remarkable absorption line spectrum of the distant radio galaxy 3C65 ($z$=1.2) obtained with the Keck 10m by Stockton et al (1995). The high signal/noise in this spectrum details the Balmer lines and Ca II doublet sufficiently clearly to establish an age of >2-3 Gyr at $z$=1.2. An exhaustive survey of this kind for luminous $z$=0.5-1 ellipticals without radio emission is possible with the new generation of telescopes and would provide tighter constraints on their likely ages than results discussed here based on broad-band colours alone.

The conclusion emerging thus far is that the *stars* that make up massive cluster ellipticals formed in a narrow time interval prior to $z \simeq 3-5$. It is, of course, conceivable that there was some subsequent merging of sub-units to create these galaxies but the associated star formation cannot have been significant at recent times as this would conflict with Bower et al's $U-V$ scatter. Kauffman (1995) addresses this point directly and finds the likely merger rate in hierarchical CDM models can be reconciled with Bower et al's data if most of the merging occurred between 0.5< $z$ <1. However, the small scatter now observed in more distant clusters must presumably push any merging to yet higher redshifts.

A final caveat is that the results quoted above might only apply to the rather atypical giant ellipticals in dense cluster regions. Certainly, there is growing evidence that lower luminosity ellipticals and those in lower density environs have suffered more recent star formation (Rose et al 1994). However, as we will see in §5, there is no evidence for significant evolution in the numbers or luminosities of distant field ellipticals either.

# 3 Slow Evolution of Massive Disk Galaxies

Bergeron & Boissé (1991) and, more recently, Steidel & Dickinson (1995) have demonstrated the value of using the galaxies responsible for the metallic absorption line systems in the spectra of distant QSOs as probes of field galaxy evolution to $z \simeq 1$. Steidel & Dickinson's comprehensive survey of >50 Mg II absorbing galaxies with $\overline{z} \simeq 0.7$ indicates that the typical absorber is a massive disk galaxy whose rest-frame $B$ and $K$ luminosities exceeds $\simeq 0.06 L^*$ drawn from a luminosity function (LF) which is remarkably similar to the bright end of that observed locally (although their normalisation is somewhat higher, see §5). The colours of the absorbers likewise span the range seen locally (after allowing for redshift effects). Overall, the survey is consistent with little or no evolution for the absorbing population.

Although these results have received much attention, the sample remains small, particularly in each colour class. The absence of strong luminosity evolution from $z=0$ to 0.7 can only be claimed with confidence for the entire population. Additionally, one might worry that, by selecting galaxies in absorption, only some subset of the population that has 'settled down' is found, with the consequence that a no evolution result is guaranteed. The QSO surveys have certainly found some galaxies which do *not* give rise to absorption lines (so-called 'interlopers'), but many of these are blue star-forming dwarfs (Steidel et al 1993). That the absorbers and interlopers are non-overlapping photometric samples is perhaps the most interesting result.

The Mg II surveys discussed above demand high resolution QSO absorption line spectroscopy, deep imaging in the vicinity of the QSO and subsequent spectroscopy of candidate galaxies. Without the final stage to confirm a galaxy has precisely the same redshift as that of the Mg II absorption line, no firm association can be made. At higher redshift, this becomes harder both because the galaxies are fainter, and because there is a dearth of suitable features for estimating redshifts with optical spectrographs for $z > 1.3$.

For the purposes of deriving an absorber LF, Aragón-Salamanca et al (1994) showed that a statistical association will suffice and may be obtained from imaging alone provided (i) the field counts are known in the chosen imaging passband to the limits explored and (ii) the only excess galaxies close to the QSO sightline represent the absorbing population. The latter is a moot point given the possible physical association of galaxies with QSOs but this point could be checked by imaging a control sample of QSOs selected without the appropriate absorption lines. Aragón-Salamanca et al examined the nature of sources producing clustered CIV absorption lines via deep $K$-band imaging of $\simeq 11$ QSOs whose absorbers have $1.5 < z < 2.5$. Their $K$-band LF is similar to that observed locally implying no dramatic luminosity changes to quite high redshifts. However, there is considerable uncertainty in matching the absolute LF normalisations since the excess expected from their choice of clustered CIV lines is difficult to estimate. Steidel & Dickinson have used similar arguments to extend their original Mg II sample beyond $z \simeq 1$ and find equivalent results.

A related development is the search for galaxies producing damped Lyman alpha absorption. These absorbers have a high HI column density and a frequency suggesting they occur in the disks of high redshift galaxies (Wolfe et al 1987). Can these sources be imaged and, if so, what is their luminosity and colour? As the near-IR $k$-correction remains modest to quite high $z$, deep $K$ band images might detect starlight from these disks, particularly if they are luminous systems like their counterpart metallic absorbers at lower redshift. Their detection would also provide some useful constraints on the amount of dust in the high $z$ Universe. A small impact parameter from the QSO sight line is expected if they originate in galaxy disks of high HI column density and so exquisite conditions are important if the galaxy is to be seen next to or 'underneath' the very bright QSO.

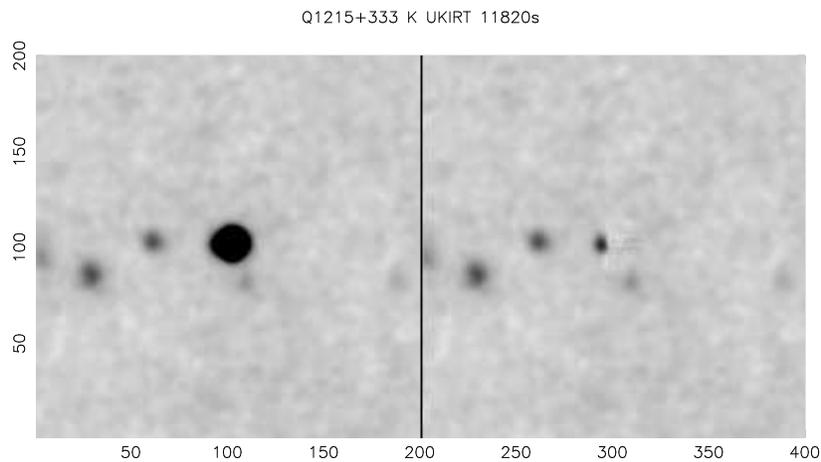

*Figure 3: The subtraction of the $K \simeq 16$ QSO Q1215+333 from a deep UKIRT image reveals a promising absorber candidate with $K=20.2$ which may be responsible for the damped Ly$\alpha$ absorption line at $z_{abs}=2.001$ (Aragón-Salamanca et al 1995).*

Figure 3 shows a deep $K$ image taken from the damped Ly$\alpha$ identification programme of Aragón-Salamanca et al (1995). For 10 suitable QSOs with $z_{abs} \simeq 2$-2.5, several faint candidates have been identified close to the QSO but none implies luminosities much brighter than $L^*$. HST imaging of such candidates will be particularly useful in clarifying the scale lengths of the star forming component and in constraining the possibility of dust extinction via imaging at shorter rest-wavelengths. When NICMOS becomes available on HST, routine surveys for such galaxies can be contemplated and reliable LFs constructed to quite high $z$.

Suggestive though these results are, the high redshift absorber programmes need to overcome several problems. Firstly, only for the $z <1.3$ Mg II samples do we have the vital spectroscopic confirmation. Spectroscopic work is difficult and does not benefit from multi-object techniques that have accelerated other studies. However, with high-throughput infrared spectrographs it may ultimately be possible to secure the redshifts of the distant Mg II, CIV and Ly$\alpha$ absorbers. It seems this will only be feasible if the absorbers are star-forming systems with strong emission lines. Another problem is that we may only be seeing a subset of the population at any redshift via these techniques. Part of the galaxy population does not participate in the absorption progress and determining that 'interloper' fraction becomes harder as we proceed to higher redshifts. We appear to have been rather lucky in this respect with the MgII absorbers, but this may not be the case for the Ly$\alpha$ systems (Lanzetta et al 1995).

## 4 Redshift Surveys and the Dwarf-dominated Universe

The absence of any significant changes in the bright end of the normal galaxy LF discussed in §2-3 is difficult to reconcile with the puzzling excess in the field galaxy number counts (Ellis 1993). Elementary considerations suggest the large surface density seen in the deepest work (Metcalfe et al 1995) implies a dwarf-dominated Universe (Cowie & Lilly 1990). How can this result be consistent with the observed LF today which apparently does not contain a huge population of dwarfs? Either we are seriously undercounting our local Universe (Phillipps & Driver 1995) or there were many more dwarf galaxies in the recent past. The clue to separating these two possibilities lies in securing redshifts for large numbers of faint galaxies. If the local LF has been underestimated, we expect to uncover large numbers of nearby dwarfs.

It might seem something of an embarrassment to admit that the local LF remains so uncertain. However, the evidence for a serious error is only circumstantial. Efstathiou et al (1988) analysed 326 galaxies in 5 Schmidt-sized fields to $b_J \simeq 17$ and found the Schechter (1976) form appropriate. They determined uncorrected Schechter parameters $< M_B^*, \alpha, \Phi^* >$ of <-19.7, -1.07, 0.0156> ($H_o$=100 kms sec$^{-1}$ Mpc$^{-1}$). The more extensive panoramic sparse-sampled APM-Stromlo southern survey of 1769 galaxies (Loveday et al 1992) found similar parameters <-19.7, -1.11, 0.0140>. However, both surveys constrain the faint end slope only to $M_{bJ} \simeq$-16. An upturn at fainter luminosities such as that claimed for the Virgo cluster (Binggeli et al 1988) cannot be formally ruled out.

Importantly, a local population of feeble sources would have an Euclidean count slope dominating the faint counts and diminishing any evolution that would otherwise be inferred (Kron 1982). Furthermore, surveys limited at relatively bright apparent magnitudes adopt high surface brightness detection thresholds and may be poorly-suited for finding intrinsically faint galaxies (McGaugh 1994). Clearly, the most reliable constraints on the faint end of the local LF comes from spectroscopy at those faint limits where the contribution can be directly measured (Glazebrook et al 1995a).

A related uncertainty which has plagued the subject concerns the question of the *absolute* normalisation of the LF. Although the Efstathiou et al and APM-Stromlo surveys have consistent values of $\Phi^*$, the number counts steepen beyond their apparent magnitude limits, $17 < b_J < 20$, more than can be accounted for by non-evolving models, suggesting southern volumes with $b_J < 17$ may be unrepresentative (c.f. Maddox et al 1990).

Combining 'benchmark' $b_J < 17$ redshift surveys and deep equivalents within narrow faint apparent magnitude slices is not ideal. Broadhurst et al, Colless et al & Glazebrook et al were only able to compare the faint redshift distribution $N(z)$ with empirical predictions based on the bright survey. At no redshift was there a sufficient range in luminosity to examine the form of the LF directly. Moreover, the number of pencil beams was small ($\simeq 5$ each) raising the worry that clustering may affect the conclusions.

My colleagues and I have therefore compiled a new 'Autofib' survey (using the robotic fibre positioner built for the Anglo-Australian Telescope – Parry and Sharples 1988) spanning a wide apparent magnitude range in many directions and enabling a direct reconstruction of the LF at various redshifts. It includes $\simeq 700$ redshifts from earlier magnitude-limited surveys and $\simeq 1000$ new redshifts within $17 < b_J < 22$ (see Table 1). Further details of this new survey are contained in Heyl (1994) and preliminary results have been presented in Colless (1995); the full analysis will appear in Ellis et al (1995).

In estimating absolute magnitudes of faint galaxies, account must be taken of the $k$-correction which depends on the galaxy's (unknown) spectral energy distribution (SED). At a given redshift, $k_{b_J}(z)$ changes by $\simeq 1$ magnitude across the Hubble sequence (c.f. King & Ellis 1985). Heyl (1994) devised a classifier based on cross-correlation of the faint spectra against the wide aperture local spectral catalogue of Kennicutt (1992). Knowing the Kennicutt class which best matches the faint galaxy, the $k$-correction is determined with reference to King & Ellis' compilation. Realistic simulations suggest the correct spectral class is returned to within $\pm 1$ class for 90% of the cases; 6 classes span the entire sequence.

Table 1: The Autofib Redshift Survey

| Survey | $b_j$ limits | Area deg$^2$ | Fields | Redshifts |
|---|---|---|---|---|
| DARS(Peterson et al 1986) | 11.5–16.8 | 70.80 | 5 | 326 |
| BES(Broadhurst et al 1988) | 20.0–21.5 | 0.50 | 5 | 188 |
| LDSS-1(Colless et al 1990,1993) | 21.0–22.5 | 0.12 | 6 | 100 |
| Autofib bright | 17.0–20.0 | 5.50 | 16 | 480 |
| Autofib faint | 19.5–22.0 | 4.70 | 32 | 546 |
| LDSS-2(Glazebrook et al 1995a) | 22.5–24.0 | 0.07 | 5 | 73 |
| TOTAL | | | | 1713 |

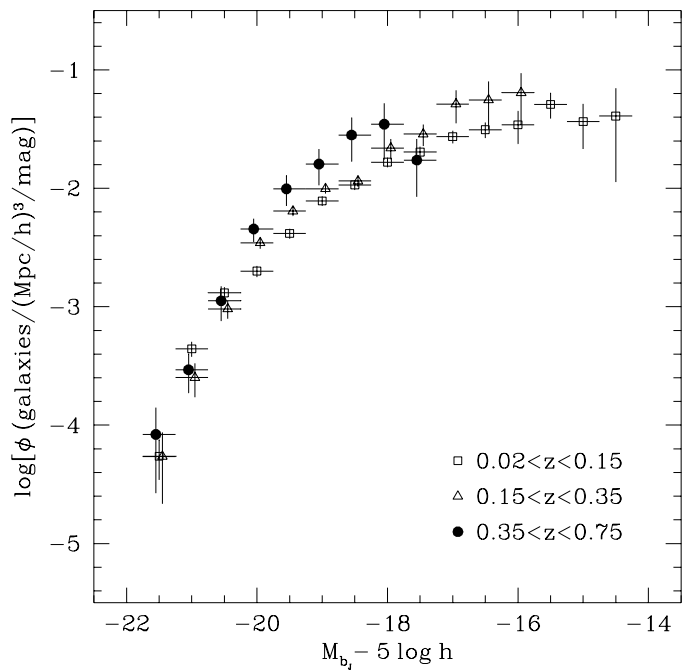

Figure 4: Luminosity functions derived from the Autofib survey (Ellis et al 1995) in different redshift intervals.

The large number of faint pencil beams in the Autofib survey leads to new constraints on the *local* LF as well as on its form at high $z$. 560 galaxies in the survey have $0 < z < 0.1$ but few are less luminous than $M_{bJ} \simeq -16.0$; most with $b_J > 22$ galaxies lie beyond $z \simeq 0.1$. The paucity of low luminosity galaxies severely limits the size of any possible upturn in the local LF to $M_{bJ} \simeq -14$. As the photometric data used to select these galaxies penetrates to surface brightness limits below $\mu_{bJ}$=26.5 arcsec$^{-2}$, it becomes hard to argue that the flat LF derived in earlier work arises from selection biases. The redshift distribution at $b_J$=24 eliminates the possibility that the faint source counts are significantly contaminated by a population of low luminosity galaxies under-represented in the original $b_J$ <17 surveys (Glazebrook et al 1995a) and gives further support to the higher $\Phi^*$ normalisation discussed in §3.

So where does the excess count come from? Figure 4 shows a highly suggestive steepening of the faint end slope of the LF with increasing redshift. Formally, a change in *shape* with $z$ is significant at the 99.9% level. However, as in §3, there is no obvious brightening in the bright end of the LF to $z \simeq 1$. As before, it seems the LF is composed of two components – a luminous part evolving very slowly, if at all, over $z$ <1-2 *plus* a rapidly-evolving sub-$L^*$ component whose fate remains unclear.

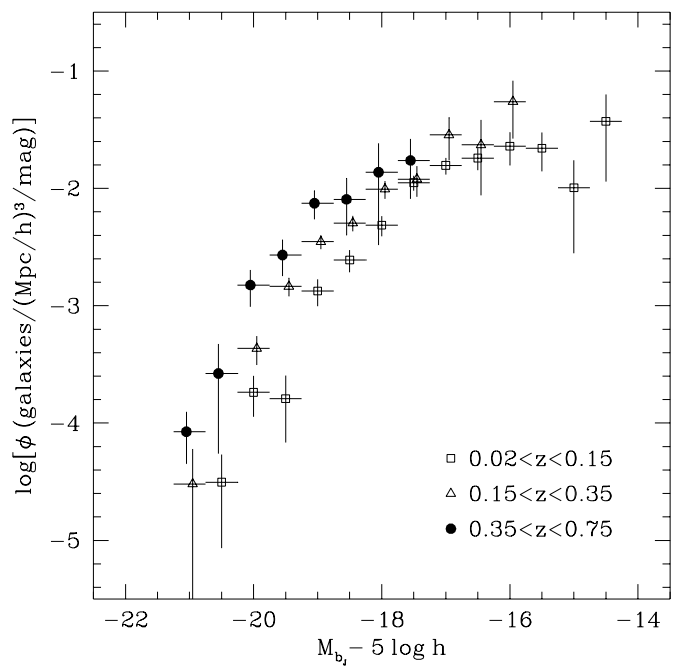

*Figure 5: Luminosity functions at various redshifts from the Autofib survey for those galaxies whose rest-frame [O II] 3727 Å equivalent width exceeds 20 Å. The evolutionary trends for this class are sufficient to explain those seen in the overall population presented in Fig. 4.*

What distinguishes the galaxies that lie in these two components? The missing clue appears to be related to the star-formation rate, as originally suggested by Broadhurst et al (1988). Figure 5 shows how the galaxies with the strongest [O II] emission dominate the evolutionary trends. The luminosity density con-tributed by this class of star-forming galaxies has dropped by a factor $\simeq 10$ since $z \simeq 0.5$. Significantly, at the present epoch, such systems all lie at the faint end of the LF, whereas at modest redshifts they occupy a wide range of luminosities.

## 5 Faint Galaxy Morphologies from HST

The repair of HST is adding a new dimension to the study of faint field galaxies. Early Cycle 4 images (Griffiths et al 1994) demonstrated the ability to recognise resolved morphological features in $I \simeq 22$ galaxies the bulk of which lie at $0.4 < z < 0.7$ (Lilly 1993). A new set of questions can be addressed with such data: (1) What is the morphological mixture of the high $z$ field population? (2) Are the faint blue galaxies a distinct morphological population?

Ultimately a large sample of HST morphologies *with redshifts* is required. However, meanwhile, a considerable amount can be determined from images alone. The largest collection is currently provided by the 'Medium Deep Survey' (PI: R. Griffiths). In this key project, WFPC-2 is used in parallel mode according to primary pointings defined by other observers. Redshifts for the galaxies surveyed have to be secured later. An increasing amount of HST imaging is being done in the reverse mode: i.e. primary WFPC-2 imaging of larger fields with existing multi-slit spectroscopy. First results of these programmes are also emerging (Schade et al 1995, Cowie et al 1995, Broadhurst, priv. comm.).

The most significant claim to date has been made by the Medium Deep Survey. Over 300 $I < 22$ galaxies were classified on a simple E/S0: Spiral: Irr/Pec scheme by Glazebrook et al (1995b) and the number-magnitude counts determined as a function of type. Counts for an enlarged sample of 550 galaxies are shown in Figure 6. The results illustrate that the spiral and early-type classes show little evidence for evolution. Significantly, both classes fit the predictions only if the absolute normalisation is $\times \simeq 2$ higher than that derived from the 17th magnitude surveys (c.f. §2). On the other hand, the irregular and peculiar galaxies show a count slope much steeper than expected, consistent with significant evolution. Although the overlap with the spectral samples remains small, the limited data suggests that the [O II]-strong sources which decline dramatically in number since $z \simeq 1$ *are* the morphologically unusual examples in the HST samples. A morphologically-distinct population of star forming sources appears to be responsible for the well-established excess population of faint blue galaxies.

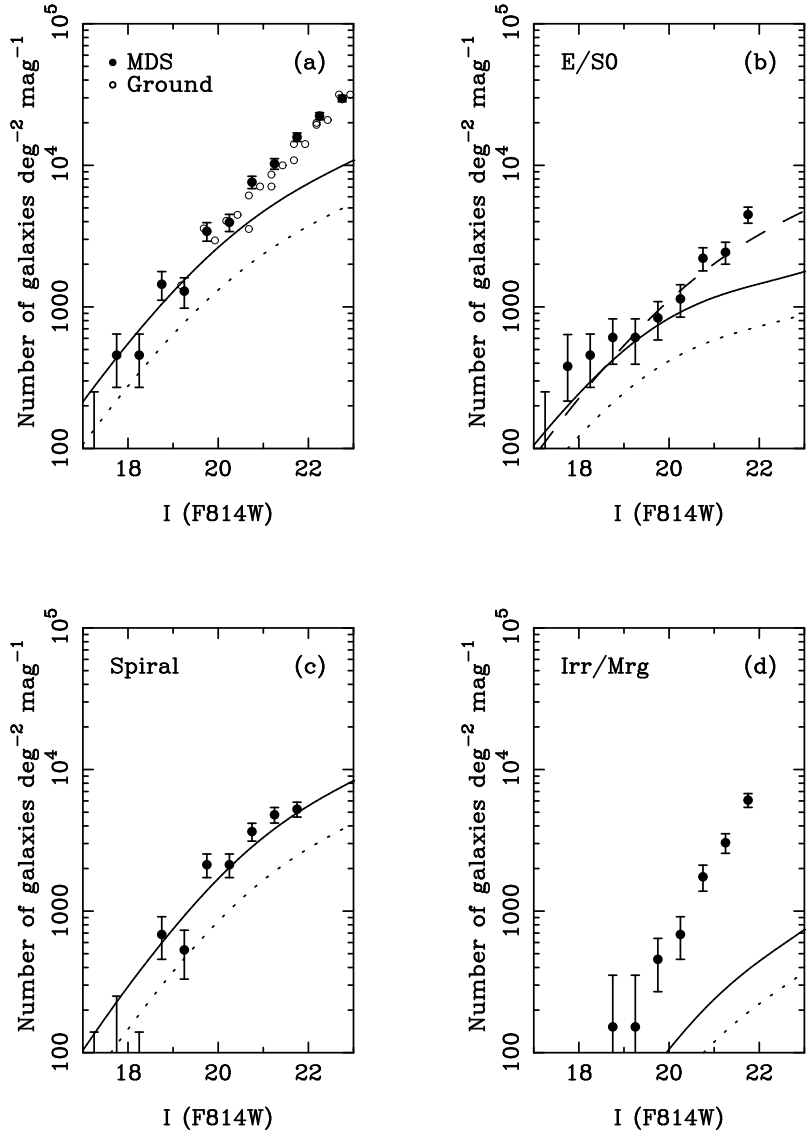

Figure 6: Morphological counts of 550 galaxies with $I < 22$ from the HST Medium Deep Survey. The sample is enlarged from that presented by Glazebrook et al (1995b). Counts of 'regular' Hubble types are consistent with no evolution (solid and dotted lines depending on $\Phi^*$) but those of irregular/peculiars show a steeper slope suggesting rapid evolution.

But can we really recognise galaxy morphologies readily at these faint limits? Is there not a danger that a faint galaxy whose signal to noise is poor in some respect or other is too readily cast into the 'peculiar/irregular' category? To quantify this possibility, Figure 7 shows a set of regular Hubble sequence galaxies observed (a) in the $B$-band at $z \simeq 0$ and (b) as simulated in the $I$-band viewed by HST at $z=0.8$. $k$-corrections are allowed for via the positioning of these two passbands and corrections have been applied to allow for the reduced HST night sky background and various other instrumental effects.

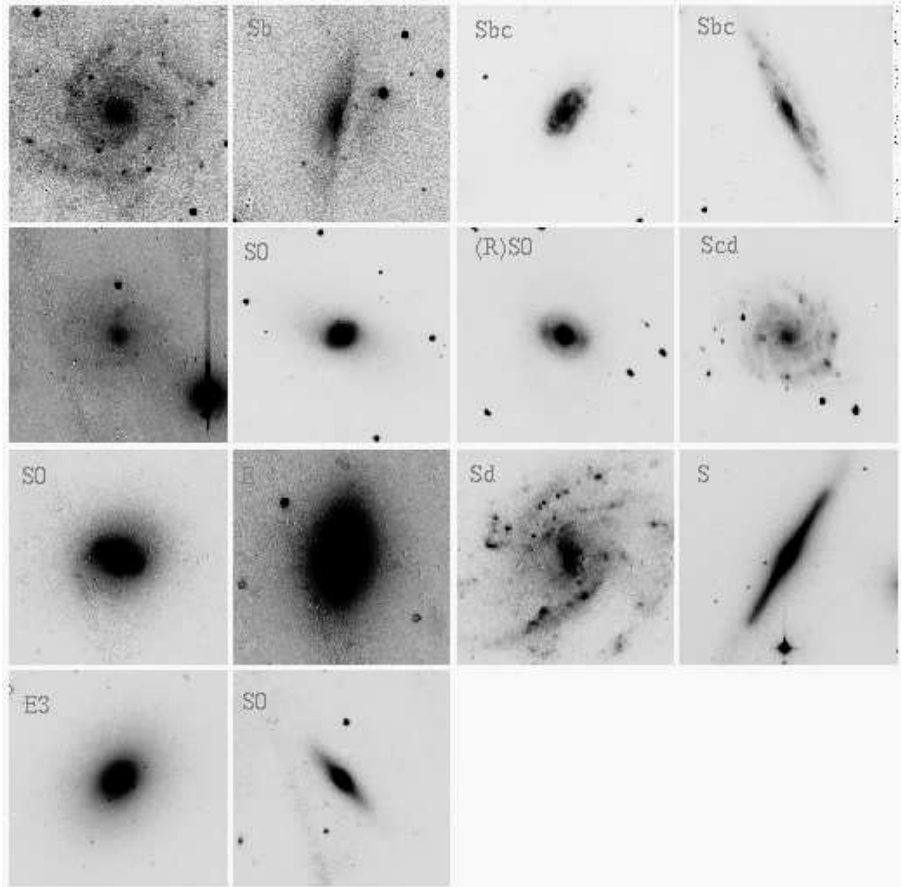

*Figure 7(a): A selection of regular Hubble sequence galaxies viewed at $z=0$ in the B band (courtesy of M. Pierce)*

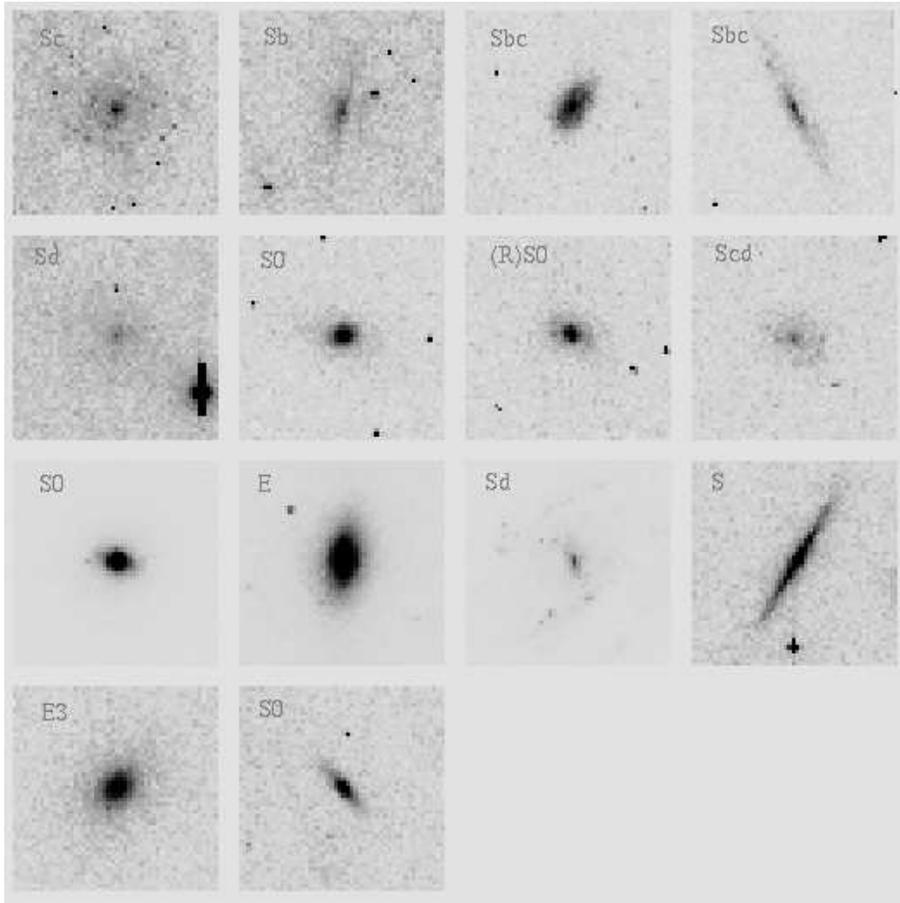

*Figure 7(b): As (a) simulated for WFPC-2 in the I band for the same sample viewed at z=0.8 taking into account various instrumental and background effects (see text for details).*

Such simulations suggest the trends seen in the MDS are reasonably robust although certain flocculent late type spirals could conceivably be misinterpreted as interacting/peculiar systems/ Extending these trends, one might worry that star forming galaxies might ultimately break up into disparate H II regions of small apparent size (c.f the 'train-wrecks' in Dickinson's image of 3C324 at $z=1.2$ and Dressler et al's (1993) 'nascent' galaxies supposedly at $z \simeq 2$).

# 6   Conclusions

Several different lines of enquiry converge on the identification of two very different galaxy populations. Massive regular galaxies (roughly defined as sym-metrical spirals and ellipticals more luminous than $\simeq 0.1$ L*) appear to have changed little since $z \simeq 1$-2 and the bulk of the star formation in the earlier types occurred very early indeed ($z >3$-5). In contrast, even the modest redshift ($z \simeq 0.5$) Universe was dwarf dominated and this dwarf population has rapidly declined in recent times. These systems are less well-formed and show intense star formation (possibly of a bursting type).

I see two unsolved problems. In hierarchical theories of galaxy formation such as those that form the basis of simulations by Kauffman et al (1993) and Cole et al (1994), dwarfs and globular clusters form early in great abundance around dark matter halos with low circular velocities. Merging reduces the number of dwarfs with time but steep faint end slopes to the local LFs are still expected today unless feedback and other processes are very effective. The end point of these simulations is closer to the observed Universe at $z \simeq 0.5$ (Figure 4) rather than $z=0$. A further difficulty is the older ages for the massive galaxies which is, at first sight, counter to the expectations of hierarchical growth.

The clue to solving this problem, as recognised by Babul & Rees (1992), lies in *preventing* a major component of the dwarf population from forming stars until late times. Babul & Ferguson (1995) show how star formation might be inhibited by the ionising UV background until recent epochs. A feature of this picture would be the sudden collapse of such systems over a narrow redshift interval when the associated gas neutralises (Efstathiou 1995). In this respect, the irregular morphologies of the HST images and the wide range in luminosity and redshift at which the evolving population is seen (Fig. 5) is puzzling. Although it is claimed the small physical sizes of the MDS sources (Im et al 1995) is consistent with this picture, not enough work has been done to separate the evolving and quiescent components, or in understanding obvious biases which may render large sources as compact ones (Figure 7).

The second unsolved problem concerns the fate of these star-forming dwarfs. Figures 4-5 demonstrate this problem observationally very clearly. Excess objects in the high $z$ LF could fade to lower luminosities, but given the wide redshift range over which the [OII]-strong sources are seen, the decay would appear to be too rapid for conventional stellar mass functions. As most of the present-day low luminosity systems are gas-rich (McGaugh 1995), no conclusive observational evidence for fading has yet been presented.

Alternatively, the dwarfs could disappear via minor merging into larger galaxies. Broadhurst et al (1992) showed qualitatively how self-similar merging is consistent with both the optical and infrared photometric and redshift data. The picture is largely under attack from theorists who object to the high merging rates required and the likely effect this will have on the morphologies of the merger products. Observationally, it is exceedingly difficult to test the idea further. Although the angular correlation function, $w(\theta)$, shows no convincing statistical excess of pairs on small scales, a significant percentage (10-20%) of the MDS sample appear to be interacting systems supporting earlier investigations (Zepf & Koo 1989). Given the rapid timescale involved, even if only a fraction of these are genuinely merging, the phenomenon must be significant.

None of the proposed solutions to either unsolved problem is particularly attractive. On the positive side, the disparity between the two populations of galaxies revealed by a number of independent observations is a sound result which will hopefully lead to the understanding of a major new physical process in galaxy formation. On the negative side, I do worry about the absurdity of switching on a dominant optical component of the Universe in the most recent few Gyr only to allow it to fade rapidly into obscurity. This seems specifically contrived to fit the puzzling observations! However, "when you have eliminated the impossible, whatever remains, however improbable, must be the truth" (Conan-Doyle 1920).

**Acknowledgements:** The *Autofib* survey involves Matthew Colless, Tom Broadhurst, Jeremy Heyl and Karl Glazebrook and the Medium Deep Survey is led by Richard Griffiths. The 'Morphs' project involves Harvey Butcher, Warrick Couch, Alan Dressler, Gus Oemler, Ray Sharples and Ian Smail. I thank all co-workers and particularly Alfonso Aragòn, Karl Glazebrook and Ian Smail for their generosity in allowing me to present data prior publication. I acknowledge financial support from the IAS and an upgrade from British Airways.